\documentclass[a4paper,fleqn,usenatbib]{mnras}

\usepackage{mathptmx}

\usepackage[T1]{fontenc}
\usepackage{ae,aecompl}
\usepackage{amsmath}
\usepackage{mathrsfs}

\usepackage{graphicx}	
\usepackage{amsmath}	
\usepackage{amssymb}	
\usepackage{indentfirst}

\def\l{\left}
\def\r{\right}
\def\f{\frac}

\def\hub{{\mathcal H}}

\title[IC in EFT]{A note on the initial conditions within the effective field theory approach of cosmic acceleration}

\author[X. Liu et al.]{
Xue-Wen Liu,$^{2,3}$\thanks{E-mail: liuxuewen14@itp.ac.cn}
Bin Hu,$^{1}$ and Yi Zhang$^{4}$\\
$^{1}$Department of Astronomy, Beijing Normal University, Beijing, 100875, China\\
$^{2}$CAS Key Laboratory of Theoretical Physics, Institute of Theoretical Physics,\\
Chinese Academy of Sciences, P.O. Box 2735, Beijing 100190, China \\
$^{3}$School of Physical Sciences, University of Chinese Academy of Sciences,
No.19A Yuquan Road, Beijing 100049, China\\
$^{4}$College of  Science, Chongqing University
of Posts and Telecommunications, Chongqing 400065, China
}

\date{Accepted XXX. Received YYY; in original form ZZZ}

\pubyear{2017}

\begin{document}
\label{firstpage}
\pagerange{\pageref{firstpage}--\pageref{lastpage}}
\maketitle

\begin{abstract}
By using the effective field theory approach, we investigate the role of initial condition for the dark energy or modified gravity models. In details, we consider the constant and linear parametrization of the effective Newton constant models. Firstly, under the adiabatic assumption, the correction from the extra scalar degree of freedom in the beyond $\Lambda$CDM model is found to be negligible. The dominant ingredient in this setup is the primordial curvature perturbation originated from inflation mechanism, and the energy budget of the matter components is not very crucial. Secondly,  the iso-curvature perturbation sourced by the extra scalar field is studied. For the constant and linear model of the effective Newton constant, there is no such kind of scalar mode exist. For the quadratic model, there is a non-trivial one. However, the amplitude of the scalar field is damped away very fast on all scales. Consequently, it could not support a reasonable structure formation. Finally,  we study the importance of the setup of the scalar field starting time. By setting different turn-on time, namely $a=10^{-2} $ and $a=10^{-7} $, we compare the cosmic microwave background radiation temperature, lensing deflection angle auto-correlation function as well as the matter power spectrum in the constant and linear model. We find there is an order of $\mathcal{O}(1\%)$ difference in the observable spectra for constant model, while for the linear model, it is smaller than $\mathcal{O}(0.1\%)$.
\end{abstract}




\section{Introduction}

On large scales of our universe, the cosmic structure density inhomogeneity is much smaller than unity. Hence, inhomogeneity growth could be well described by the cosmic linear perturbation theory. For this dynamical system, there are two essential ingredients, namely gravitational instability and initial conditions. For the gravitational instability, the gravity force acts on all the matter components in the universe and drives matters to form cosmic structure. And the initial conditions come from the mechanism in the extreme early universe, such as the cosmic inflation epoch. In the concordance model of cosmology, the gravity is described by the General Relativity (GR) and the primordial Initial Condition (IC) is modeled by a spatial power law parametrization which physically comes from the curvature perturbation in  single field inflation paradigm.

One of central problems in the  modern cosmology is cosmic acceleration, \emph{i.e.} the phase of accelerated expansion recently entered by the Universe~\cite{Riess:1998cb, Perlmutter:1998np}.
Now we are still lack of a satisfactory theoretical explanation for this.
Within the framework of General Relativity, an accelerated expansion phase could be achieved by adding an extra ingredient in the energy budget, commonly referred to as Dark Energy (DE). It could be static, as in the cosmological constant model ($\Lambda$CDM), or dynamical.  Alternatively, one could also construct models which modify the laws of gravity on the large scales in order to achieve self-accelerating solutions in the presence of negligible matter which is called `modified gravity' (MG).

With the ongoing and upcoming surveys, such as Planck~\footnote{\url{http://sci.esa.int/planck}}, SDSS~\footnote{\url{http://www.sdss.org}}, DES~\footnote{\url{http://www.darkenergysurvey.org}}, LSST~\footnote{\url{http://www.lsst.org}} and Euclid~\footnote{\url{http://sci.esa.int/euclid}}, we shall anticipate a wealth of high precision large scale structure data. Hence, it is crucial to test the laws of gravity via a model independent method against the evolution of linear cosmological perturbations. In this work we apply the effective field theory (EFT) approach to the phenomenon of cosmic acceleration~\cite{Gubitosi:2012hu,Bloomfield:2012ff}.
Based on a parametrized action, EFT offers a unified and model-independent framework for performing agnostic tests of gravity and accurately exploring most of the viable models of cosmic acceleration.
Any model of DE/MG which introduces a single extra scalar field and allows for a well defined Jordan frame, could be mapped exactly into the EFT language, without the need of  approximation. In a further leap towards observations, this framework has been implemented into the linear Einstein-Boltzmann solver \texttt{CAMB/CosmoMC}~\footnote{\url{http://camb.info}}~\cite{Lewis:1999bs,Lewis:2002ah},  resulting in the powerful patches \texttt{EFTCAMB/EFTCosmoMC} which are introduced in~\cite{Hu:2013twa,Raveri:2014cka,Hu:2014oga,Hu:2014sea,Hu:2015rva} and   publicly available~\footnote{\url{http://www.eftcamb.org}}. A similar code, namely \texttt{hi\_class}~\cite{Bellini:2014fua,Zumalacarregui:2016pph} based on \texttt{class}~\cite{Blas:2011rf}, is also publicly released~\footnote{\url{http://www.thphys.uni-heidelberg.de/~zumalacarregui/hi_class.html}}.

In this paper, we apply the \texttt{EFTCAMB/EFTCosmoMC} code performing the following calculation. The \texttt{EFTCAMB/EFTCosmoMC} fully exploit the unifying nature of the EFT formalism, and provide a highly accurate and efficient setup to test gravity on the linear cosmological scales. Model independent parametrizations as well as specific models of DE/MG could be explored in \texttt{EFTCAMB/EFTCosmoMC}, without the need of specializing the set of perturbed equations to the case under study. The equations in \texttt{EFTCAMB/EFTCosmoMC} are indeed implemented in the EFT language, without the use of any quasi-static approximation. Several models of DE/MG are already implemented in the publicly available version of \texttt{EFTCAMB}, including $f(R)$ gravity~\cite{Hu:2016zrh}, Ho\v rava gravity~\cite{Frusciante:2015maa}, \emph{etc}.

According to their ``thermal dynamical'' properties, the initial conditions could be classified into two categories, namely adiabatic and iso-curvature/entropy modes. For the former, it is based on the assumption that, the relative abundances of different particle species were determined directly from the Lagrangian describing local physics, one would expect those abundance ratios to be spatially constant because all regions of the universe would share an identical early history, independent of the long wavelength perturbations. The entropy of the universe is conserved and the primordial curvature perturbation setup the initial conditions. In contrast, if in one or more matter components  particles were created or annihilated,  the entropy modes are able to dominate the early-time evolution which  are also dubbed as iso-curvature  mode. As the source of  entropy mode varies, they could be further specified as CDM iso-curvature, baryon iso-curvature, neutrino velocity iso-curvature modes~\cite{Ma:1995ey,Perrotta:1998vf,Bucher:1999re}, \emph{etc}.
Recent observations, such as Planck~\cite{Ade:2015lrj}, favor the  adiabatic mode  produced by the single field inflation model.

In this paper, we are aiming to understand the role of extra scalar field in the DE/MG model in the determination of initial conditions. We try to answer two questions: how to set the adiabatic mode and  is there exist an extra scalar field iso-curvature mode. The rest of the paper is organized in the following ways. In  section \ref{sec:IC},  the algorithm of deriving both the adiabatic and iso-curvature modes in the EFT formalism with corresponding model setup will be present. In section \ref{sec:res}, the CMB and matter spectra results is shown. Finally, the conclusion is drawn in section \ref{sec:con}.

\section{Initial Conditions in EFT approach}\label{sec:IC}
In this section, we investigate the IC in  DE/MG models, including both adiabatic and iso-curvature modes, within  the constant and linear $\Omega$ model of the pure EFT parametrization \cite{Hu:2013twa,Hu:2015rva}.

\subsection{Model Setup and Numerical Algorithm}
The EFT action for the background operators in the unitary gauge reads
\begin{align}\label{action_Stuck}
  S=\int\sqrt{-g}& \left\{\frac{m_0^2}{2}[1+\Omega(\tau)]R+\Lambda(\tau)-a^2c(\tau)\delta g^{00}+\cdots\right\} + S_{m},
\end{align}
where $m_0$ is the Planck mass and the sub-script ``$m$'' stands for all the rest regular cosmic components. The scalar field hidden in the gauge field could be reproduced by a time-shift, namely St$\ddot{u} $ckelberg trick:~$\tau\to\tau+\pi(\bold{x})$. In the EFT language, we could find that the only operators which are involved into the background evolution are, $\Omega(\tau)$, $c(\tau)$ and $\Lambda(\tau)$. Hence, we get the following two relationships with the background Hubble parameter and its first time derivative. It means that, for a fixed background evolution, such as the $\Lambda$CDM one, only one of them could be chosen as an extra free function in the model parametrization
\begin{align}
&\hub^2=\f{a^2}{3m_0^2(1+\Omega)}(\rho_m+2c-\Lambda)-\hub\f{\dot{\Omega}}{1+\Omega}, \\
\label{acc1}
&\dot{\hub}=-\f{a^2}{6m_0^2(1+\Omega)}\l(\rho_m+3P_m\r)-\f{a^2(c+\Lambda)}{3m_0^2(1+\Omega)}-\frac{\ddot{\Omega}}{2(1+\Omega)},
\end{align}
where the dot represents the derivative with respect to the conformal time.

As a widely used cosmological model, the non-minimally coupled gravity models could be simply parametrized by the cosmological effective Newton constant or $\Omega$ function\footnote{$G_{\rm eff} = (1+ \Omega(\tau))G_{\rm N}$ with $8\pi G_{\rm N}=1/m_0^2$.}. In this paper, we consider the constant $\Omega$ models (Model 1)
\begin{align}
\Omega(\tau)=\Omega_0,
\end{align}
and the linear $\Omega$ models in term of the scale factor $a$ (Model 2)
\begin{align}
\Omega(\tau) = \Omega_0a\;.
\end{align}

As pointed out in \cite{Hu:2015rva}, the constant $\Omega$ model is not just a simple redefinition of the gravitational constant. The requirement of having a $\Lambda$CDM background with a non-vanishing $\Omega$, that would change the expansion history,  which indicates  a scalar field is sourced  to compensate this change. The scalar field could  then interact with the other matter fields and modify the behavior of cosmological perturbations, consequently change the CMB power spectra and the growth of structure. For instance, in the constant $\Omega$ model, $c(\tau)$, which is vanishing in general relativity, is non-zero and reads  $c=\Omega(\rho_m+P_m)/2$. More interestingly, this model shares some similarity with the extended Brans-Dicke model \cite{Lima:2015xia}, namely the deviation of the effective Newton constant on the cosmological scale from those on the earth, is not vanishing in the radiation epoch, when we assign the ICs. This hints that the modification to gravity is not a late-time phenomena, it originate from early epoch.

For the linear $\Omega$ models, it is in the contrary. Since $\Omega$ decays along the arrow of back of time, the linear $\Omega$ model approximate General Relativity in the early time. So, by definition, these two models capture  different modifications of  gravity essentially. In the following, we exam the influence of the modification of  ICs on the observables.

Firstly, we list the dynamical equations of the cosmic fluids, the metric field as well as the extra scalar field on the super-horizon regime in the radiation epoch when  setting up the ICs of the Einstein-Boltzmann solver.
At this epoch, photons ($\gamma$) and baryons ($b$) are tightly coupled with each other via the Thomson scattering process, and forms a single plasma with $\theta_b=\theta_{\gamma}$, where $(\delta_i,\theta_i,\sigma_i)$ denote for the density contrast, velocity potential and anisotropic energy stress of the ``$i$'' component
\begin{align}
\label{eq:gamma}
&\dot{\delta}_\gamma + {4\over 3}\theta_\gamma + {2\over 3}\dot{h}=0\,,
	\qquad \dot{\theta}_\gamma - {1\over 4}k^2 \delta_\gamma = 0, \\
 \label{eq:baryon}
&  \dot{\delta_{b}}+\theta_{b}+\frac{1}{2}\dot{h}=0,
  ~~~~\dot{\theta_{b}}+\frac{\dot{a}}{a}\theta_{b}=0.
 \end{align}
The massless neutrinos ($\nu$) behave similar to photons except that they could develop the anisotropic stress
\begin{align}
&\dot{\delta}_\nu + {4\over 3}\theta_\nu + {2\over 3}\dot{h}=0,
   	\qquad \dot{\theta}_\nu - {1\over 4}k^2
	(\delta_\nu-4\sigma_\nu) = 0 \,,\nonumber\\
  & \dot{\sigma}_\nu - {2\over 15}(2\theta_\nu+\dot{h}+6\dot{\eta}) = 0.
\end{align}
 The cold dark matter ($c$) only gravitationally interacts with other cosmological components and is treated as pressure-less perfect fluid. Usually we set our reference co-moving with cold dark matter
 \begin{align}
 \label{eq:cdm}
  \dot{\delta_{c}}=-\frac{1}{2}\dot{h},
  ~~~~ \theta_{c}=0,
 \end{align}
 where the $\eta$ and $h$ are the metric scalar perturbations in the synchronous gauge.
 Finally, the modified Einstein equations are 
  \begin{align}
  \label{eq:ee1}
  &\dot{\eta}=\frac{2a^2\sum\rho_{m}\theta_{m}}{3k^2(1+\Omega)}+\frac{1}{H_0(1+\Omega)}\{(2a^2\cdot c-\hub\dot{\Omega}+\ddot{\Omega})\pi+\dot{\Omega}\dot{\pi}\},  \, \\
  \label{eq:ee2}
  &\dot{h}=\frac{2}{G}(\frac{k^2\eta}{\hub}+\frac{a^2\sum\rho_m\delta_m}{2\hub(1+\Omega)}+\frac{L}{H_0}), ~~G =1+\frac{\dot{\Omega}}{2\hub(1+\Omega)}\,, \\
  &L =\frac{1}{2\hub(1+\Omega)}\{\dot{\Omega}\pi(3\dot{\hub}-k^2)-3\hub\dot{\Omega}\pi-3\hub\dot{\Omega}\pi \nonumber\\&~~~~~+2(\dot{\pi}-2\hub\pi)a^2\cdot c   \, \}. 
  \end{align}
and the Klein-Gorden equation for the $\pi$ field reads
  \begin{align}
  \label{eq:KG}
  &A = a^2\cdot c + \frac{3\dot{\Omega}^2}{4(1+\Omega)},~~B=a^2 \cdot \dot{c}+4a \dot{a}\cdot c+\frac{\dot{\Omega}}{4(1+\Omega)}(4a^2\cdot c  \nonumber \\
  &~~+ 6\hub\dot{\Omega} +6\ddot{\Omega}+3\frac{\ddot{a}}{\dot{a}}\dot{\Omega}), ~~C=a^2\hub \cdot\dot{c}+2a^2(3\hub^2-\dot{\hub})\cdot c \nonumber\\
  &~~+\frac{3}{2}(\ddot{\hub}-2\hub^3)\dot{\Omega},~~D=a^2\cdot c+\frac{3\dot{\Omega}}{4(1+\Omega)}, \nonumber\\
  &E=(\frac{a^2\cdot c}{2}+\frac{3\dot{\Omega}^2}{8(1+\Omega)})\dot{h}-\frac{a^2\dot{\Omega}}{4(1+\Omega)}\rho_m\delta_m  \nonumber\\
  &A\ddot{\pi}+B\dot{\pi}+C \pi+k^2\cdot D \pi + H_0 \cdot E = 0.
  \end{align}
Up to now, we complete the dynamical quantities
\begin{align}
\label{eq:var}
\delta_{\gamma},~\delta_{\nu},~\delta_{b},~\delta_{c},~\theta_{\gamma},~\theta_{\nu},~\theta_{b},~\theta_c,~\sigma_{\nu},~\pi,~h,~\eta
\end{align}
and their equations (\ref{eq:gamma}-\ref{eq:KG}).

The next step is to set up their initial values and solve the equations. We Taylor expand each of the above physical quantities in terms of $k\tau$, which is much smaller than unity in the radiation epoch.
It will turn the above linear differential equation set into a series of linear algebraic equation set. And we ask the coefficients at each order of $k\tau$ vanish.

Following this procedure, we reduce our task into solving a linear algebraic equation problem with multiple degrees of freedom. However, physically, we have only \emph{single} parameter per ``$k$'' mode from the early universe, such as the amplitude of the conserved curvature perturbation from inflation. In order to bridge the gap between them, we need to specify some constraint relationship among the dynamical quantities, which are classified in terms of ``adiabatic'' or ``iso-curvature'' perturbation modes.  We will discuss these in the following sub-section.

\subsection{Adiabatic mode}
For global observers, such as the one co-moves with the matter fluid, the universe what he/she observed is a perturbed one with respect to the FLRW solution.
However, in the vicinity of the small patches of the universe, one could always be able to find a local observer, who feels living in a non-perturbed FLRW background. The differences among the local observers are the local expansion rate $\mathcal{H}(\tau_{\rm loc})$ and spatial curvature $\mathcal{K}(\tau_{\rm loc})$.
This statement is always true, as long as the Birhkoff theorem is valid, no matter what kind of initial conditions are.

Moreover, as pointed out in Baumann's lecture note\footnote{Cosmology, Part III Mathematical Tripos, \url{http://www.damtp.cam.ac.uk/user/db275/Cosmology/Lectures.pdf}}, if the perturbation is adiabatic, such as the one predicted from the single field inflation model, the global perturbations have the property that local state of matter at some spacetime point $(\tau,\bold{x})$ of the perturbed universe is the same as the background universe at some slightly different time $\tau+\delta\tau(\bold x)$. Hence, in another word, the difference in the cosmic fluid at different places is nothing but the time delay
\begin{align}
\delta\rho_m(\tau,\bold{x})=\bar{\rho}_m(\tau+\delta\tau(\bold{x}))-\bar\rho_m(\tau)=\dot{\bar\rho}_m\delta\tau(\bold{x})\;.
\end{align}
The adiabatic initial condition tells us the time delay is the same for all the matter components.
This is the standard picture in the $\Lambda$CDM case.

Furthermore, in the DE/MG models, there exists an extra dynamical scalar degree of freedom $\phi(\tau,\bold{x})$.
Within the EFT language, the perturbed scalar field could also be absorbed by a time shift $\tau\to\tau+\pi(\bold{x})$, which  is the
St$\ddot{u} $ckelberg field
$\delta\phi(\tau,\bold{x})=\bar{\phi}(\tau+\pi(\bold{x}))-\bar{\phi}(\tau)=\dot{\bar{\phi}}\pi(\bold{x})$.
So, we could simply generalize the above adiabatic statement including the extra scalar field
\begin{align}\label{pi_adiab}
\delta\tau=\frac{\delta\rho_i}{\dot{\bar\rho}_i}=\frac{\delta\phi}{\dot{\bar\phi}}=\pi.
\end{align}
The equation (\ref{pi_adiab}) is the major result in this sub-section. It tells us that there is a constraint relationship among the matter components and EFT $\pi$ field.

As argued above, for an adiabatic observer, to the lowest order of $k\tau$ (its zeroth order), he/she could only feels the gravitational potential well, namely the conserved curvature perturbation on the super-horizon regime
\begin{align}
\label{eq:var_adiab}
\eta_0=1\;,\delta_{\gamma0}=\delta_{\nu0}=\cdots=\pi_0=0,
\end{align}
where the sub-script ``0'' denote for the expansion order in $k\tau$. Here we normalize the curvature perturbation into unity.
And equation (\ref{pi_adiab}) sets up the rules how the matter components and $\pi$ field fall into the potential well.

\subsection{$\pi$ field iso-curvature mode}
In this sub-section, we consider the possible alternative scenario to the adiabatic IC mode. This means that the primordial seed of the structure formation is not the conserved curvature perturbation from inflation.  The curvature perturbation ($\dot h, \eta$) during the radiation epoch is not the dominant component to drive the structure growth. To the lowest order of $k\tau$ (its zeroth order), we could set $\dot h=\eta=0$ at the initial time, namely ``iso-curvature'' mode.

From the Einstein equation (\ref{eq:ee1}, \ref{eq:ee2}), we could conclude that, in the $\Lambda$CDM scenario ($\pi$ vanishes), neither the energy density nor velocity perturbations of the dominant components (radiation=photon+neutrino) in this epoch contribute to the metric perturbation. So, we could allow the density or velocity perturbations of the sub-dominant components (baryon and CDM) to seed the structure formation. So, eq. (\ref{eq:baryon},\ref{eq:cdm}) are the driven term of the system, namely ``baryon'' or ``CDM'' density iso-curvature modes. Besides of these, we notice that in the radiation epoch, it is the sum of the photon and neutrino density/velocity perturbations contribute to the metric perturbation. Hence, there is a space for a fine tuning among these two components, which could cancel their contributions between each other, namely ``neutrino'' iso-curvature density/velocity mode.
For details, we recommend the readers the classical piece of works~\cite{Perrotta:1998vf,Bucher:1999re}.

However, for the DE/MG model, mathematically, there is a possibility that the $\pi$ field equation (\ref{eq:KG}) could dominate the system.
Similar to the adiabatic mode, to the zeroth order of $k\tau$, we set $\pi_0=1$ and all the rest of quantities in (\ref{eq:var}) vanish
\begin{align}
\label{eq:var_adiab}
\pi_0=1\;,\delta_{\gamma0}=\delta_{\nu0}=\cdots=\eta_0=0.
\end{align}

\section{Results}\label{sec:res}

\subsection{Adiabatic mode}
Combined with Einstein, fluid and Klein-Gorden equations, the constant $\Omega$ model gives
\begin{align}
\label{eq:ic1}
&\delta_{\gamma}=C\frac{3+3\Omega_0}{3+\Omega_0}(k\tau)^2,~~\delta_{b}=\delta_{c}=\frac{3}{4}\delta_{\nu}=\frac{3}{4}\delta_{\gamma},
\\&\theta_{c}=0,~\theta_{\gamma}=\theta_{b}=C\frac{1+\Omega_0}{4(3+\Omega_0)}k^4\tau^{3},
\\&\theta_{\nu}=C\frac{(1+\Omega_0)(23+4R_{\nu}+39\Omega_0)}{4(3+\Omega_0)(15+4R_{\nu}+15\Omega_0)}k^4\tau^3,
\\&\sigma_{\nu}=-C\frac{6(1+\Omega_0)(1+3\Omega_0)}{(3+\Omega_0)(15+4R_{\nu}+15\Omega_0)}(k\tau)^2,
\\&\pi=-CH_0\frac{3+3\Omega_0}{4(3+\Omega_0)}k^2\tau^3,
\\&\eta=-3C+\frac{3(5+4R_{\nu}-15\Omega_0)(1+\Omega_0)}{4(4R_{\nu}+15+15\Omega_0)(3+\Omega_0)}(k\tau)^2,
\end{align}
where $R_{\nu}$ is the neutrino fraction in the radiation epoch. 
And the linear $\Omega$ model gives
\begin{align}
&\delta_{\gamma} =
 C k^2 \tau^2 ,
   ~~\delta_{b}=\delta_{c}=\frac{3}{4}\delta_{\nu}=\frac{3}{4}\delta_{\gamma},
\\&\theta_{c}=0,~~\theta_{\gamma} =\theta_{b}=
C \frac{1}{12}k^4 \tau^3,~~
\theta_{\nu} =
 \frac{k^4 \tau^3C (23 +
      4R_{\nu})}{12 (15 + 4R_{\nu})}
\\&\sigma_{\nu}= -\frac{2Ck^2\tau^2}{15 + 4R_{\nu}}-
 2Ck^2 \tau^3\frac{(35+12R_{\nu})H_0\Omega_0\sqrt{\Omega_r}}{(15 + 2R_{\nu}) (15 + 4R_{\nu})}
\\&\pi =
 -\frac{1}{4}C H_{0}k^2 \tau^3,~~
h = -\frac{3}{2}Ck^2 \tau^2,
\\&\eta = -3C+
  k^2 \tau^2C\frac{5+
      4R_{\nu}}{4 (15 + 4R_{\nu})} -
 5C k^2 \tau^3\frac{H_{0}\Omega_0\sqrt{\Omega_r} (35+12R_{\nu})}{2 (15 + 2R_{\nu}) (15 + 4R_{\nu})},
 \label{eq:ic2}
\end{align}
where $C$ is   a integration constant  of order $10^{-5}$ which could be calculated from the amplitude of curvature perturbation from inflation.
Equations (\ref{eq:ic1}-\ref{eq:ic2}) show the modification to the standard matter components and metric perturbations, start from the sub-leading order terms.
The leading order term is fixed via the adiabatic evolution request (\ref{pi_adiab}). As for the $\pi$ field, its IC start from the order of $H_0k^2\tau^3$, which is also sub-leading compared with the
conserved curvature perturbation from inflation, namely $3C$ term in $\eta$.

\begin{figure}
\begin{center}
  \includegraphics[width=0.5\textwidth]{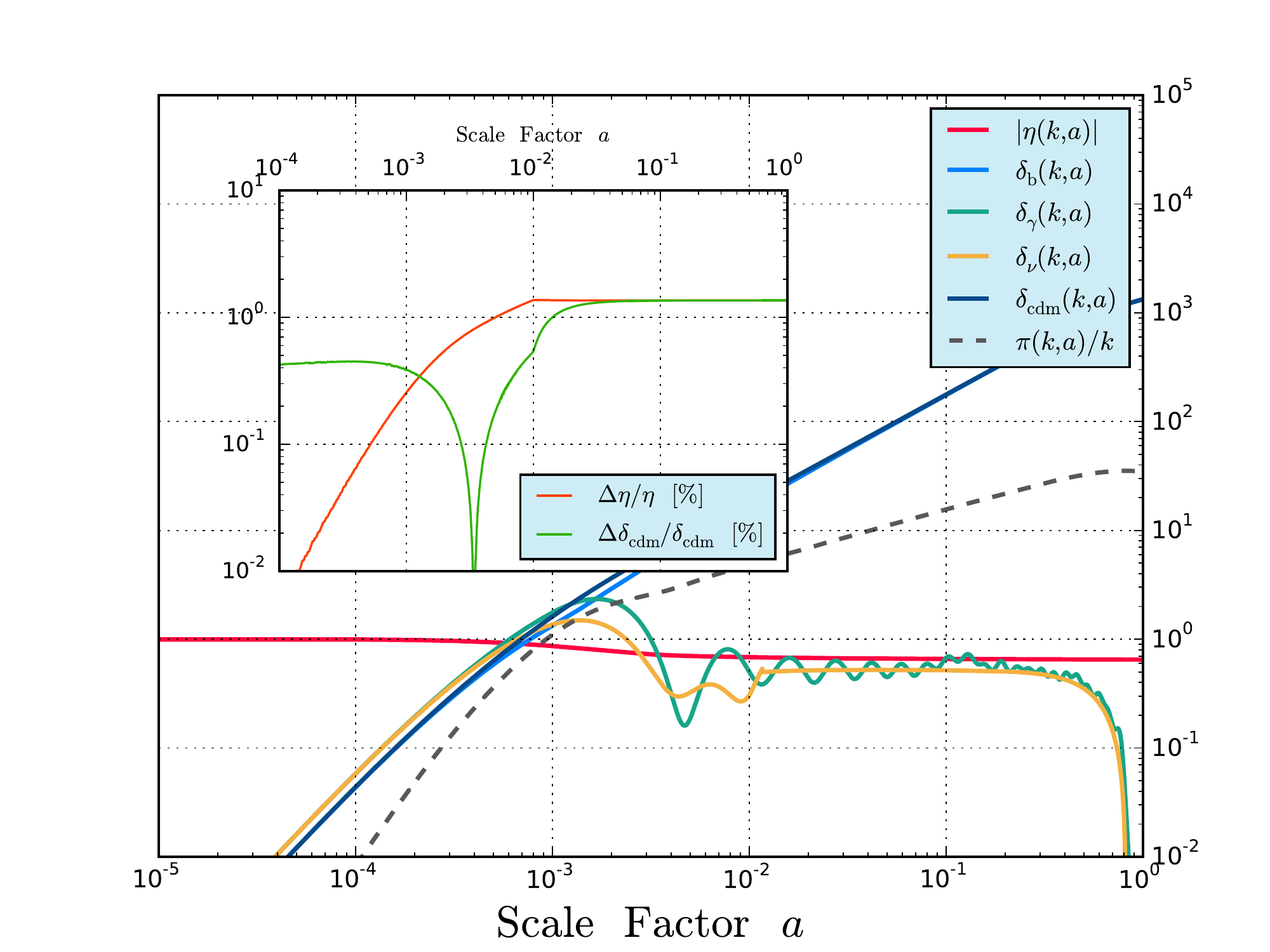}
  \caption{\label{Fig.transf}The relative difference in several dynamical variables for $k=10^{-3}\cdot{\rm Mpc}^{-1}$. Here we choose the constant $\Omega$ model with $\Omega_0=0.01$.}
\end{center}
\end{figure}
Physically,  on the super-horizon scale, a synchronous adiabatic observer feels his/her local patch as a primordial gravitational potential well plus a unperturbed single matter plasma.
As shown in the major panel of Figure. \ref{Fig.transf}, in this regime the important ingredient is the metric perturbation (the red curve), the amplitude of the matter density perturbation is much smaller
than the former. With the time pass by, the gravity will tell the matter how to fall into the primordial potential well.

The next step is to implement these analytical results into the Einstein-Boltzmann solver.
For $\Lambda$CDM cosmology, such as \texttt{CAMB} \cite{Lewis:1999bs}\footnote{http://camb.info}, we set the ICs in the deep radiation dominant era, namely $a=10^{-7}$.
However, in the current public version of \texttt{EFTCAMB}, the $\pi$ field is integrated from $a=10^{-2}$ until $a=1$. On one hand, it is valid as long as the modification of gravity is only a late-time phenomenon, such as the case in
$f(R)$ gravity \cite{Hu:2016zrh}. Via this trick, we are able to greatly reduce the cost of numerical integration. On the other hand, this algorithm becomes problematic if the model could \emph{not} recover General Relativity in the early time, such as the constant $\Omega$ model in this study. Setting the ICs of $\pi$ field at the same time as the other matter species makes the \texttt{EFTCAMB} solver more self-consistent. 

In the following, we test two different ICs. One is the $\Lambda$CDM ICs plus $(\pi=\dot\pi=0)$, the other one is obtained in equation (\ref{eq:ic1}-\ref{eq:ic2}). In both of the cases, we
start to evolve $\pi$ field since $a=10^{-7}$.Unfortunately,  \emph{we find their relative differences are negligible, and hence leaves almost no effect on the CMB angular and matter spectra.}
Again, this could be understood via  Figure. \ref{Fig.transf}. Under adiabatic assumption, the important ingredient of the cosmic dynamical system are the primordial curvature perturbation from inflation and the consequent dynamical equations, not the modification to the matter density and scalar field, which is initially sub-dominant.

Since within the EFT of DE/MG framework, we do not change the inflation epoch, so the primordial curvature perturbation is untouched.
Then, one could only get a significant effect via changing the dynamical equations. In order to demonstrate this, in the sub-panel of Figure. \ref{Fig.transf}, we show the relative differences of $\eta$ and $\delta_{\rm CDM}$ for $k=10^{-3}\cdot{\rm Mpc}^{-1}$, in the constant $\Omega$ model, one with $\Lambda$CDM evolution equation, and the other with the correct EFT equations.  We found that, initially, the density of CDM is more sensitive to the modification to the gravitational law, while the curvature perturbation $\eta$ is not. This is because the dominant term in $\eta$ during the super-horizon evolution epoch is the primordial one from inflation not the dynamics.

So, \emph{for the adiabatic mode, the important issue is not the modified initial values, but the modified equations.} Hence, for the models which significantly deviates from General Relativity in the radiation epoch, we need to evolve the $\pi$ field since $a=10^{-7}$.

\begin{figure}
\begin{center}
  \includegraphics[width=0.5\textwidth]{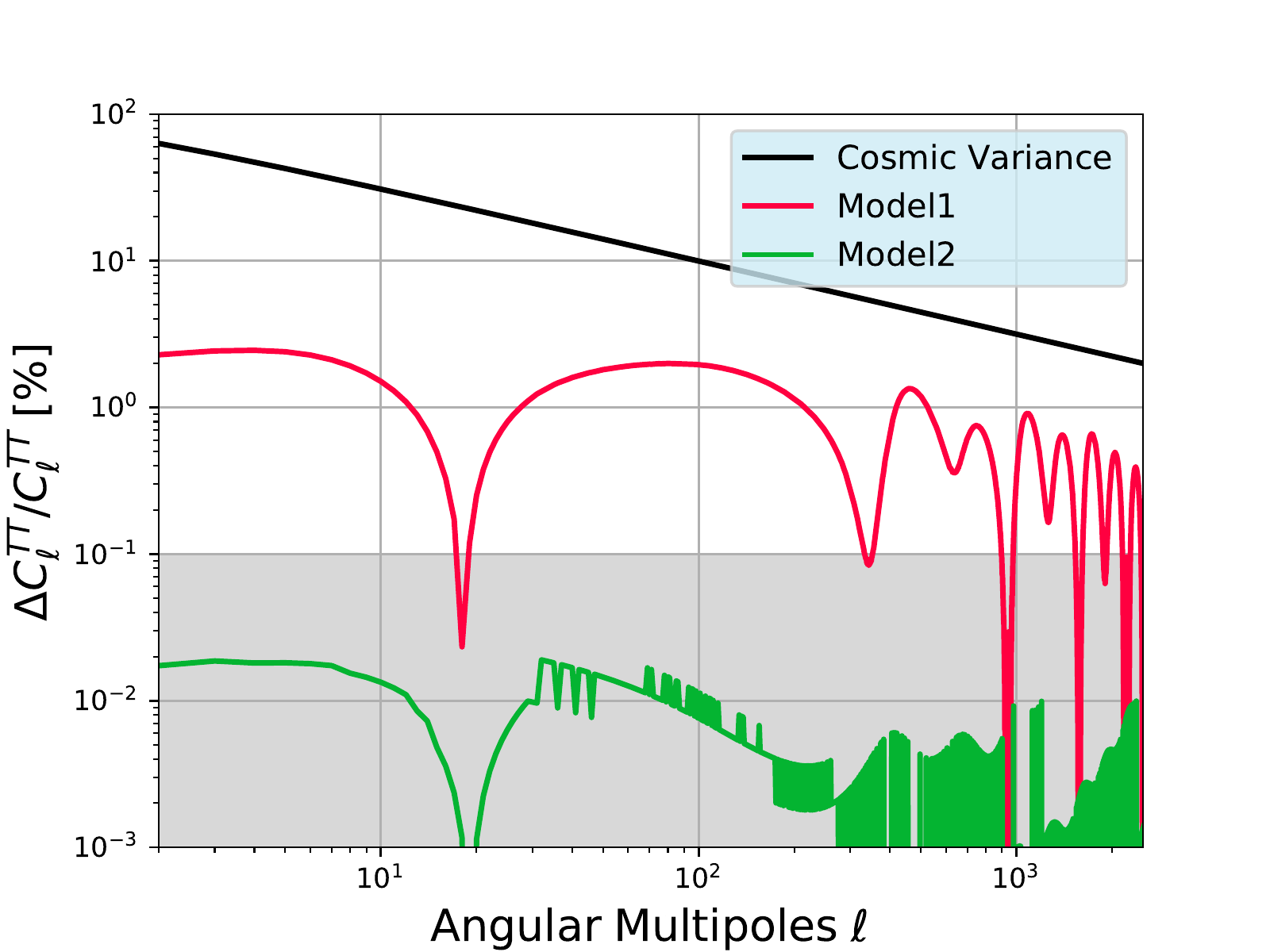}
  \caption{\label{Fig.Diff_tt} Relative difference in lensed TT spectrum between $\texttt{EFTturnonpiInitial}$= $10^{-2} $ and $10^{-7} $. Model1 :~$\Omega=\Omega_0 $~, Model2 :~$\Omega = \Omega_0\cdot a$, with $\Omega_0=0.01$. We also show the cosmic variance (black curve) to stress the importance of this difference.}
\end{center}
\end{figure}

\begin{figure}
\begin{center}
  \includegraphics[width=0.5\textwidth]{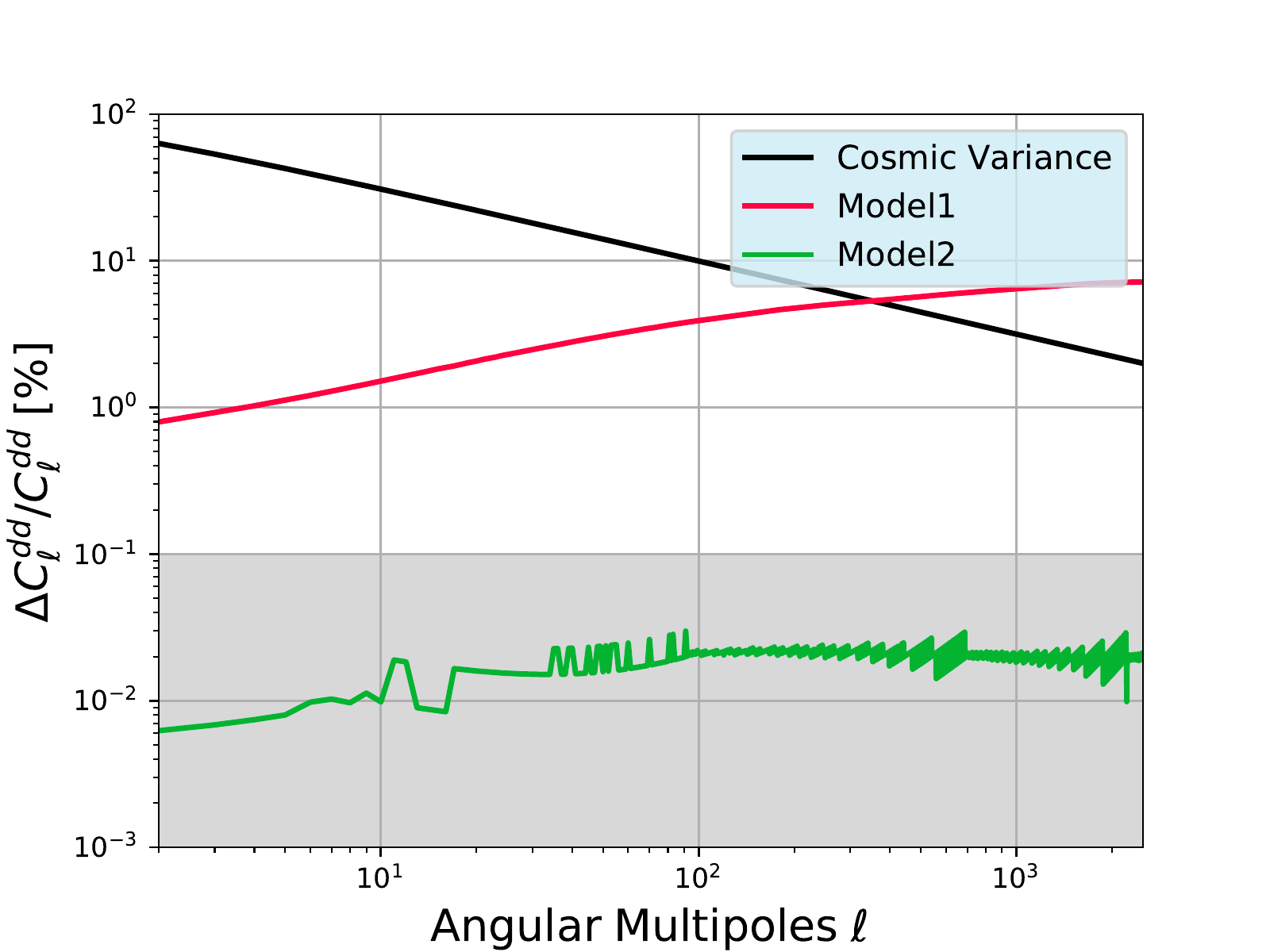}
  \caption{\label{Fig.Diff_dd} Relative difference in lensing deflection angle spectrum between \texttt{EFTturnonpiInitial}= $10^{-2} $ and $10^{-7} $. Model1 :~$\Omega=\Omega_0 $~, Model2 :~$\Omega = \Omega_0\cdot a$.}
\end{center}
\end{figure}

\begin{figure}
\begin{center}
  \includegraphics[width=0.5\textwidth]{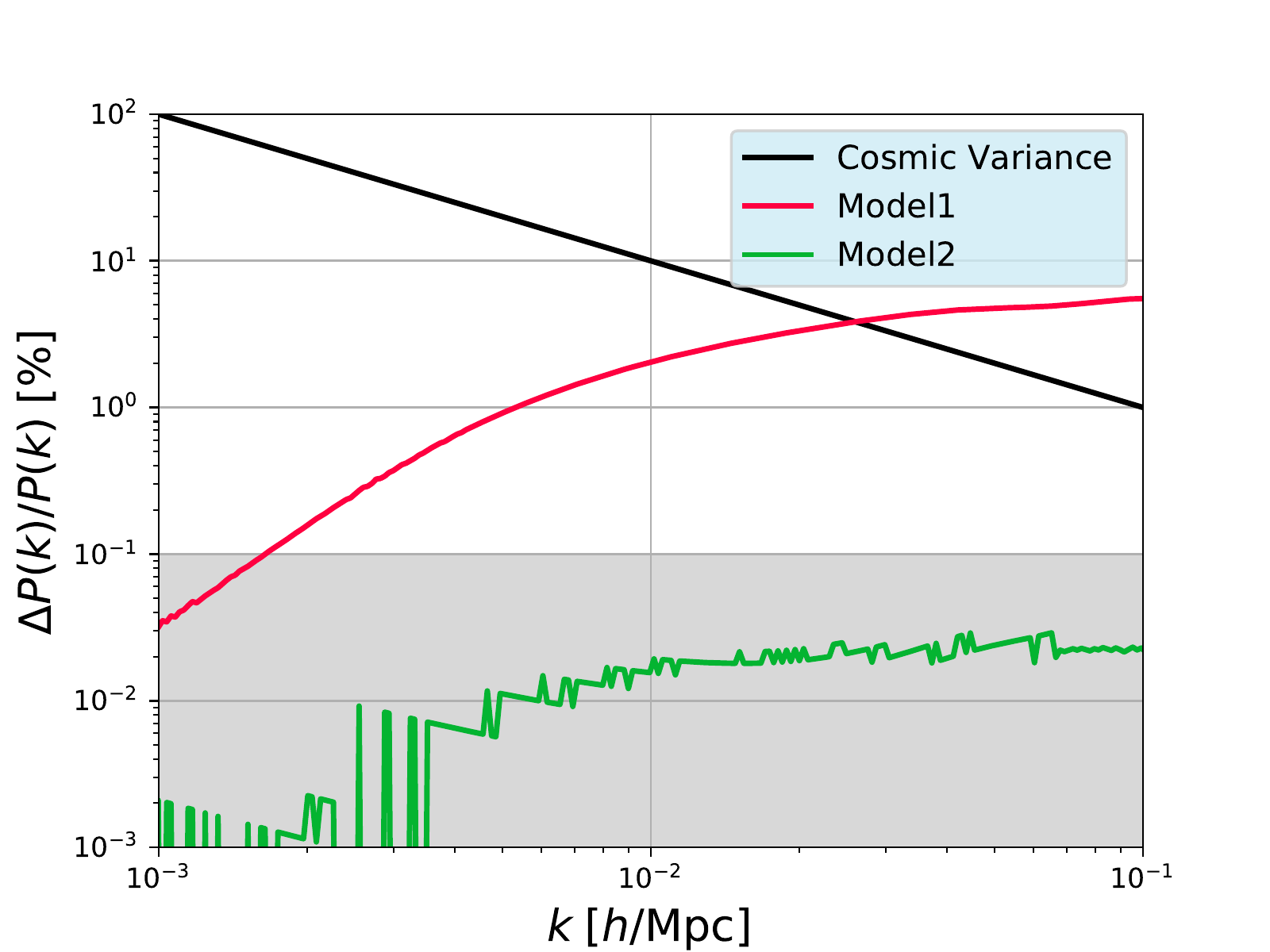}
  \caption{\label{Fig.Diff_pk} Relative difference in non-relatistic matter power spectrum between \texttt{EFTturnonpiInitial}= $10^{-2} $ and $10^{-7} $. Model1 :~$\Omega=\Omega_0 $~, Model2 :~$\Omega = \Omega_0\cdot a$.}
\end{center}
\end{figure}

Since we try to estimate the importance of the setting of $\pi$ field starting time, we calculate the CMB temperature, lensing deflection angle auto-correlation function as well as the matter power spectrum in the constant and linear $\Omega$ model, with two different $\pi$ field starting time, namely $\texttt{EFTturnonpiInitial}=10^{-2},10^{-7}$. The results are shown in Figure \ref{Fig.Diff_tt}-\ref{Fig.Diff_pk}.  In order to demonstrate its effect more significantly, we choose the $\Omega_0$ value ($0.01$) in the $2\sigma$ tail of the likelihood obtained from the combination of CMB Planck 2015~\cite{Ade:2015xua} and BAO data~\cite{Beutler:2012px,Kitaura:2015uqa,Anderson:2013oza}.

\begin{figure}
\begin{center}
  \includegraphics[width=0.5\textwidth]{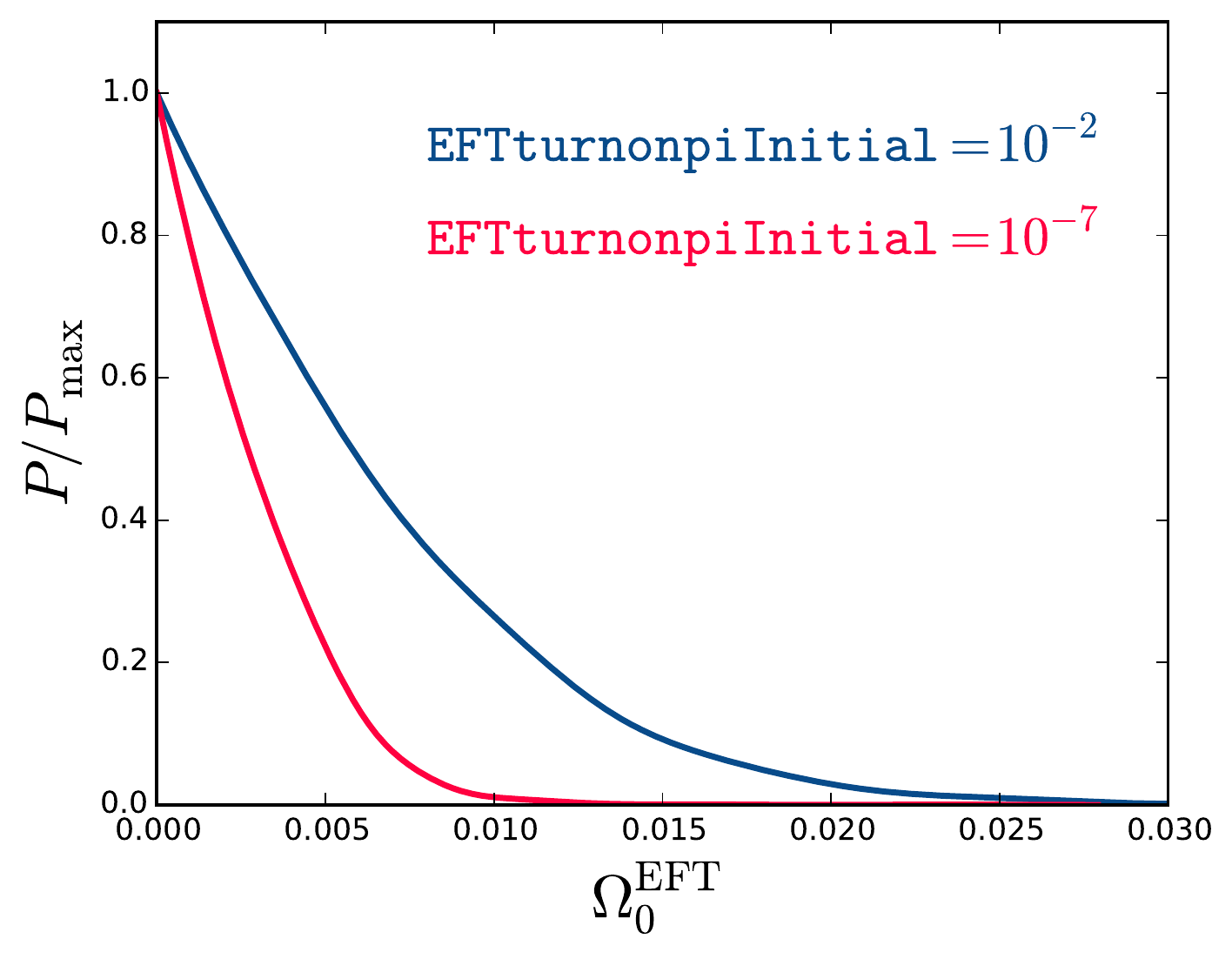}
  \caption{\label{Fig.Omega0}Posterior distribution of the effective field parameter $\Omega_0 $.}
\end{center}
\end{figure}

The fractional difference of the lensed CMB temperature spectrum is shown in Figure. \ref{Fig.Diff_tt}. We could see that for the constant $\Omega$ model, there is an order of $\mathcal{O}(1\%)$ difference, while for the linear model,
it is much smaller than $\mathcal{O}(0.1\%)$, which is close to the current data sensitivity. A similar conclusion could be drawn also from the CMB deflection angle in Figure. \ref{Fig.Diff_dd} and matter spectrum in Figure. \ref{Fig.Diff_pk}. Moreover, the modification effect seems more prominent on the smaller scales in the latter two spectra. Figure \ref{Fig.Omega0} plots the posterior distribution of MG parameter~$\Omega_0 $~in constant $\Omega$ model using the data mentioned above with \texttt{EFTCosmoMC}. It shows that we have tighter constraint on~$\Omega_0 $~in the \texttt{EFTturnonpiInitial}=$10^{-7} $ case.


\subsection{$\pi$ field iso-curvature mode}
As for the $\pi$ field iso-curvature mode, the results are quite model dependent. For the constant and linear $\Omega$ model, we do not find any solution of the
dynamical system (\ref{eq:gamma}-\ref{eq:KG}) with the initial values (\ref{eq:var_adiab}). For the quadratic $\Omega$ model, namely $\Omega(a)=\Omega_0\cdot a^2$,
there is a non-trivial solution
\begin{align}
&\delta_{b}=\delta_{c}=\frac{3}{4}\delta_{\nu}=\frac{3}{4}\delta_{\gamma}=4C\Omega_0H_0\Omega_r\tau,~\theta_{c}=0,
\\&~~\theta_{\gamma} =\theta_{b}=
 \frac{2}{3}C\Omega_0H_0\Omega_r(k\tau)^2,~\sigma_{\nu}=\frac{12}{5+4R_{\nu}}C\Omega_0H_0\Omega_r\tau,
 \\&\theta_{\nu} =
 -\frac{8(1-R_{\nu})}{3(5+4R_{\nu})}CH_0\Omega_0\Omega_r(k\tau)^2,~~\pi = C,
\\&h = -8C\Omega_0H_0\Omega_r\tau,~~\eta =\frac{(65+16R_{\nu})}{3(5+4R_{\nu})}C\Omega_0H_0\Omega_r\tau\;.
\end{align}
However, this mode could not produce any reasonable CMB angular and matter spectra. It is because the $\pi$ field behaves as a damped oscillator, whose
amplitude approaches zero very rapidly. So, the resulting spectra are several order magnitudes smaller than the adiabatic one. A similar result for the quintessence model has been reported in \cite{Perrotta:1998vf}.

\section{Conclusions}\label{sec:con}
In this paper, we aimed to build a more self-consistent algorithm to set the initial condition in the Einstein-Boltzmann solver of the dark energy/modified gravity (DE/MG) models.
As an example, we derived the adiabatic initial conditions for constant and linear $\Omega $ model within the effective field theory approach. After coding these results into the Boltzmann solver, \texttt{EFTCAMB},
we found these modifications in the initial conditions compared with those from $\Lambda$CDM are negligible on the cosmological observables.  This is due to the fact that the dominant ingredient of the density growth in the
early time is the primordial curvature perturbation on the super-horizon regime originated from inflation epoch. Compared with these primordial seeds, the modification law of gravity which tells how the matter fall into the potential
well, is not very crucial. Furthermore, we also investigate the $\pi$ field iso-curvature mode. We found there were no $\pi$ field iso-curvature modes for constant and linear $\Omega $ model, but the solution to quadratic $\Omega$ model exists. However, even in the latter case, the $\pi$ field iso-curvature mode could not produce any reasonable CMB angular and matter spectra. It is because the $\pi$ field behaves as a damped oscillator, whose amplitude vanishes very rapidly. Hence, the resulting spectra are several order magnitudes smaller than the adiabatic one.

We also discussed the importance of the setting of $\pi$ field starting time. We compared the difference of the CMB temperature, lensing deflection angle auto-correlation function as well as the matter power spectrum between the constant and linear $\Omega$ model with two different $\pi$ field starting time, namely $a=10^{-2}$ and $10^{-7}$. We found the difference were $\sim\mathcal{O}(1\%)$  in constant model and smaller than $\mathcal{O}(0.1\%)$ in linear model.
This is because, for the constant model, the modification effect of the effective Newton constant does not diminish and the model will not recover general relativity in the radiation epoch. In the contrast, the linear model does.
In another word, the differences in CMB and LSS observables just verified the fact that, the constant $\Omega$ model theoretically makes more modification to general relativity than the linear one in the early time.

\section*{Acknowledgements}
BH is supported by the Chinese National Youth Thousand Talents Program, Beijing Normal University Grant No. 312232102 and the Fundamental Research Funds for the Central Universities.
YZ is supported by CQ CSTC under grant No. cstc2015jcyjA00044 and CQ MEC under grant No. KJ1500414.



\bibliographystyle{mnras}

\bibliography{mnras_ic}




\bsp	
\label{lastpage}
\end{document}